\documentclass[aps,prb,10pt,twocolumn,letterpaper,superscriptaddress]{revtex4-1}
\usepackage{amsmath}
\usepackage{dcolumn}
\newcolumntype{d}{D{.}{.}{-1}}
\usepackage{graphicx}
\usepackage{latexsym}
\usepackage{amssymb}
\usepackage{color}

\newcommand{\rxx}{\rho_{xx}}
\newcommand{\rxy}{\rho_{xy}}
\newcommand{\rxyc}{\rho_{xy}^c}
\newcommand{\tanx}{\mbox{$a$-TaN$_{x}$}}

\begin{document}

\title{Superconductor to weak-insulator transitions in disordered Tantalum Nitride films}

\author{Nicholas P. Breznay}
\affiliation{Department of Applied Physics, Stanford University, Stanford, CA}
\altaffiliation{Present address: Department of Physics, University of California, Berkeley, Berkeley, CA}
\author{Mihir Tendulkar}
\affiliation{Department of Applied Physics, Stanford University, Stanford, CA}
\author{Li Zhang}
\affiliation{Department of Applied Physics, Stanford University, Stanford, CA}
\author{Sang-Chul Lee}
\affiliation{Department of Materials Science and Engineering, Stanford University, Stanford, CA}
\author{Aharon Kapitulnik}
\affiliation{Department of Applied Physics, Stanford University, Stanford, CA}
\affiliation{Department of Physics, Stanford University, Stanford, CA}

\date{\today}

\begin{abstract}
We study the two-dimensional superconductor-insulator transition (SIT) in thin films of tantalum nitride. At zero magnetic field, films can be disorder-tuned across the SIT by adjusting thickness and film stoichiometry; insulating films exhibit classical hopping transport. Superconducting films exhibit a magnetic field-tuned SIT, whose insulating ground state at high field appears to be a quantum-corrected metal. Scaling behavior at the field-tuned SIT shows classical percolation critical exponents $z \nu \approx$ 1.3, with a corresponding critical field $H_c \ll H_{c2}$. The Hall effect shows a crossing point near $H_c$, but with a non-universal critical value $\rho_{xy}^c$ comparable to the normal state Hall resistivity. We propose that high-carrier density metals will always exhibit this pattern of behavior at the boundary between superconducting and (trivially) insulating ground states.

\end{abstract}

\maketitle

\section{Introduction}

The common approach to understanding disordered two dimensional electron systems like amorphous thin films is to assume that as the temperature $T$ approaches zero, they can be either insulating (I) or superconducting (S). Driving them across the S-I phase boundary with disorder or magnetic field was predicted to realize a quantum phase transition --- the superconductor-insulator transition (SIT)~\cite{goldman98m}. Early experimental and theoretical studies of the SIT addressed the nature of an amplitude or phase dominated transition --- see Ref.~\onlinecite{CIQPT, *QPT} for reviews --- while more recently, evidence for unconventional ground states proximal to the SIT has emerged. These include metallic phases~\cite{mason99k, butko01a, baturina07sb} precluded by the scaling theory of localization~\cite{abrahams79al} as well as a Bosonic ``Cooper-pair insulator'' ground state~\cite{stewart_superconducting_2007, nguyen_observation_2009}. Further, the nature of the transition itself shows different regimes of behavior with changes in the degree of material disorder.

The limit of strong disorder~\cite{steiner08bk} reveals a direct, self-dual superconductor to Bosonic-insulator transition~\cite{ovadia2013, breznay_self-duality_2016}. When disorder is weak and the normal state shows quantum-corrected metallic behavior, the field-tuned SIT shows distinct critical behavior from the strong-disorder case~\cite{yazdani95k, mason99k} that is interrupted as $T\rightarrow 0$ by a still unexplained and controversial metallic state. Observed in a range of amorphous~\cite{ephron96yk,qin06vy} and crystalline~\cite{saito_metallic_2015, tsen_nature_2016} materials, a theoretical consensus has yet to emerge to explain this metallic behavior~\cite{shimshoni_transport_1998,kapitulnik01mk,spivak_quantum_2001,mulligan_composite_2016,davison_hydrodynamic_2016}; recent experiments confirm it to be of  ``failed'' superconducting character~\cite{breznay_metallic_2017}. What separates the strong- and weak-disorder behavior regimes remains unclear, and may be due to the inhomogeneous and/or granular nature of the disorder, properties of the normal state, or properties of the proximal superconductivity. As yet there are few systems where careful control of the materials system can be combined with studies across both the disorder- and field-tuned SIT.

The field- and disorder-driven suppression of superconductivity has been studied in 2D intermetallic films, such as NbSi, TiN, NbN, MoGe, and amorphous indium oxide (InO$_x$)~\cite{marrache_thickness_2008, baturina_superconductivity_2004, mondal_phase_2011, yazdani95k, sambandamurthy04ej}. Notably, TiN shows what appears to be a direct field-tuned SIT and unexpected metallic saturation in the high-field insulating state~\cite{baturina07mv}, while both NbSi and NbN can be tuned via thickness or composition across the disorder-driven SIT.~\cite{chand_phase_2012, crauste_destruction_2014} Tantalum thin films show a metal intervening in the SIT as a function of field and disorder. Qin et al.~\cite{qin06vy} studied field-tuned transitions of several nm thick Ta films, observing hysteretic I-V characteristics and metallic saturation of the resistance a low temperature similar to that observed in MoGe~\cite{mason99k} and arguing that it is not a simple heating effect~\cite{seo06qv}. In subsequent work~\cite{li10vy}, I-V characteristics were used to identify superconducting, metallic, and insulating behavior, while very recent studies have examined critical scaling in both the superconducting and metallic phases~\cite{park_scaling_2017}. In pure Ta there is no direct SIT; a metallic phase appears to intervene at all disorder strengths accessible by tuning the film thickness. These differences between materials families raise a central question: how does the SIT evolve under alternative disorder tuning parameters.

Here we study a strongly spin-orbit coupled binary metal nitride system, amorphous tantalum nitride (\tanx), where the disorder can be controlled by tuning film thickness and composition. The most disordered films show hopping transport at low temperatures, with no evidence for superconductivity. Weakly superconducting films can be driven into an insulating state beyond an applied magnetic field $H_c$, and show a temperature-independent crossing point at $H_c$ and finite-size scaling in $\rxx$ that is the hallmark of the SIT. We also report similar crossing and weak scaling behavior in the Hall effect $\rxy$, previously observed only in InO$x$~\cite{paalanen92hr}, with a non-universal value for the critical Hall resistivity $\rxyc$. Along with evidence for unusual metallic behavior on the insulating side of the field-tuned SIT, similarities with other materials point towards a universal superconductor-weak insulator transition (SWIT). Finally, as normal-state Hall effect data indicate metallic carrier concentrations for all superconducting film ($n \sim 10^{23}$ cm$^{-3}$) comparable to MoGe, Ta, and other SWIT materials, we hypothesize that such large carrier densities preclude observation of a Bosonic SIT.

This article is organized as follows: sample growth and characterization data are presented in Sec.~\ref{s:grow}, while low-temperature transport data and the disorder-tuned SIT are described in Sec.~\ref{s:expt}. The behavior in the vicinity of the field-tuned SIT is described in Sec.~\ref{s:bsit}, with scaling analyses considered in detail in Sec.~\ref{s:scaling}. Section~\ref{s:hall} describes the Hall effect in the vicinity of the field-tuned SIT, where a crossing point and faint scaling appears above the critical field $H_c$. Finally, Sec.~\ref{s:disc} summarizes the disorder and field-tuned SIT phenomenology in \tanx{} and discusses their implications.

\section{Film growth and characterization}
\label{s:grow}

We study reactively sputtered \tanx{} films, most with thickness $d\sim$5 - 20\,nm, as described in part in previous work~\cite{breznay12mt, breznay13k}. Reactive sputtering with a tantalum target in a background Ar/N$_2$ mixture is a common technique for growth of TaN films ~\cite{nie01xw,shin02kg,noda04tt}. In sputtered films, superconductivity with $T_c$ between 5 and 10\,K has been shown to be stable for a range of nitrogen stoichiometries~\cite{wakasugi97ts}; thick films can show $T_c$ values as high as 10.9~K~\cite{kilbane75,prokhorov98,wakasugi97ts,shin01gk}. Our films are grown using a commercial AJA reactive magnetron sputtering tool and pure Ta target in flowing Ar-N$_2$ gas. The Ar/N ratio controls the film Ta/N stoichiometry (and related film properties, such as the room temperature value of $\rxx$)~\cite{nie01xw,shin02kg,noda04tt}. Base pressure of the sputter deposition system was $\approx 10^{-8}$\,torr. Films were co-deposited onto several substrate materials, including intrinsic Si, Si with SiO$_2$ and Si$_3$N$_4$ buffer layers, and glass; aside from few-percent quantitative differences in film properties (such as the room temperature resistivity $\rxx$) we found no sensitivity of film parameters to the choice of substrate. All substrates were Ar-ion plasma cleaned before sputtering. Pure Ta (and Ta-rich) films were also sputtered in identical conditions and comparable behavior to previous studies~\cite{qin06vy}; results from one Ta-rich film (N3) are included below for comparison.

\begin{figure}
\centering
\includegraphics[width=1.0\columnwidth]{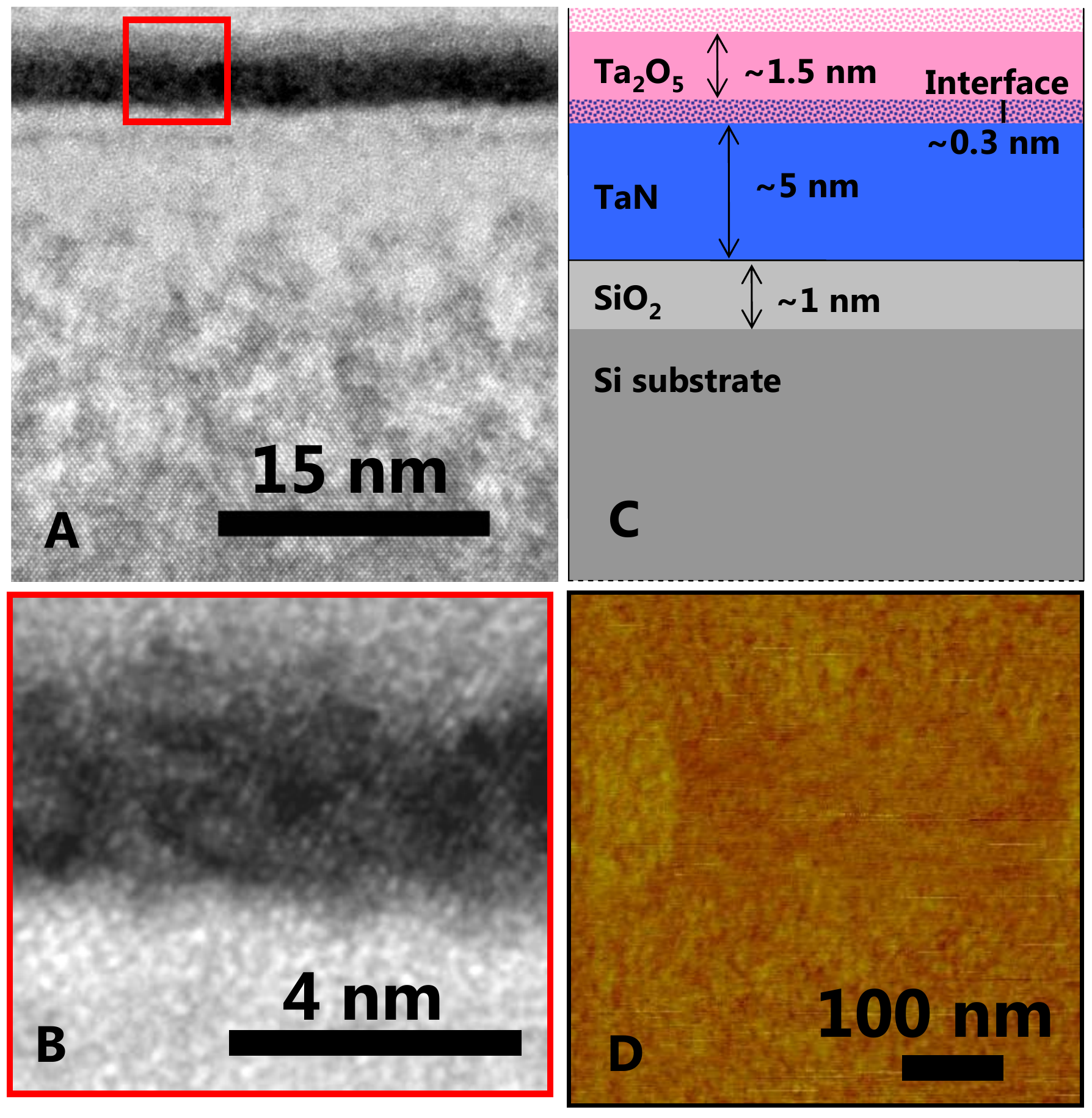}
\caption{\setlength{\baselineskip}{0.8\baselineskip} (A) Cross-sectional TEM image of \tanx{} film on Si substrate, along with (B) inset showing the $\sim 4$ nm thick amorphous film. (C) Schematic diagram of the film composition and structure based on TEM and x-ray analyses. (D) AFM image with roughness of 0.2\,nm; the full range of the color scale is 2\,nm.}
\label{f:tem}
\end{figure}

\begin{figure}
\centering
\includegraphics[width=1.0\columnwidth]{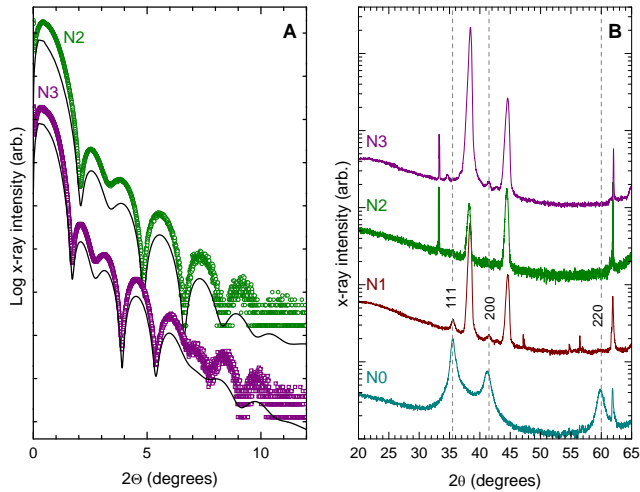}
\caption{\setlength{\baselineskip}{0.8\baselineskip} (A) X-ray reflectivity scans for samples N2 and N3, along with fits (solid lines). (B) X-ray diffraction $\theta-2\theta$ scan of samples N0-N3; curves have been vertically offset for clarity. Only sample N0 shows broad (111), (200), and (220) peaks associated with \textit{fcc}-TaN; peaks visible for films N1-N3 are due to the substrate or surface TaO$_x$ layer.}
\label{f:xrr}
\end{figure}

Several parameters can be used to tune to the disorder in \tanx{} (as reflected in e.g. $\rxx$ at room temperature), including thickness, N$_2$ partial pressure (and therefore stoichiometry $x$), and other synthesis conditions. Here we control the effective film disorder by changing the thickness and (for the $d\sim5$~nm films) nitrogen stoichiometry. Table~\ref{t:params} shows the film growth and characterization parameters, including the stoichiometry $x$, thickness $d$, and electronic properties (discussed further below).

\begin{table*}[htb]
\centering
\caption{Disordered TaN$_x$ film electronic and structural properties, including the thickness $d$, superconducting transition temperature $T_c$, longitudinal sheet resistivity $\rxx$, low-field Hall coefficient $R_H$, and carrier density $n$.}
\begin{ruledtabular}
\begin{tabular}{cccccccc}
  Sample & SC/Ins. & TaN$_x$ & $d$ & $T_c$ & $\rxx$ (300K) & $R_H$  & $n$ \\
	Name & & stoichiometry $x$ & (nm) & (K) & ($\Omega/\square$) & ($\Omega$/T) & (10$^{22}$ cm$^{-3}$) \\
  \hline
      N0    & SC    & 1.0$\pm$0.1   	& 250   & 6.4   & 10.6  & -     & - \\
      N1    & SC    & 1.1   	& 36    & 3.7   & 507   & 0.0022 & 7.9 \\
      N2    & SC    & 1.0   	& 4.0   & 2.8   & 828   & 0.014 & 11.1 \\
      N3    & SC    & 0.1   	& 4.9   & 0.5   & 526   & 0.029 & 4.4 \\
      N4    & SC    & 1.1   	& 5.9   & 0.7   & $1.3\times10^3$ & 0.011 & 9.6 \\
      N5    & Ins.  & 1.2 	  & 5.3		& -     & $8.0\times10^3$ & -     & - \\
      N6    & Ins.  & 1.4		  & 4.9		& -     & $13\times10^3$ & -     & - \\
      N7    & Ins.  & 1.6 	  & 5.2   & -     & $1.4\times10^6$ & -     & - \\
\end{tabular}
\end{ruledtabular}
\label{t:params}
\end{table*}

We characterize the film thickness, morphology, chemical composition, and homogeneity with a range of techniques. TEM measurements (Fig.~\ref{f:tem}) reveal the homogeneous and disordered \tanx{} film in sample N2 on a Si substrate; this film is 4\,nm thick and approximately stoichiometric. The \tanx{} film cross section is enlarged in Fig.~\ref{f:tem}B; the scale bar in this panel is 4\,nm. X-ray photoelectron spectroscopy (XPS) measurements confirm the film stoichiometries, with an accuracy of $\approx 10\%$, shown in Tab.~\ref{t:params}; depth profiling uncovers a $\sim$1\,nm surface oxide layer on all films. A schematic of the film cross section including a the Si substrate, \tanx{} film, along with interfacial SiO$_2$ and TaO$_x$ layers (with roughness $\sim$0.5\,nm) is shown in Fig.~\ref{f:tem}C. Scanning atomic-force microscopy measurements, shown in Fig.~\ref{f:tem}D for sample N2, show excellent homogeneity in the film thickness. The rms roughness is 0.1~nm (or 2\% of the TaN layer thickness) over a $\sim1\mu m^2$ region, with no evidence for granularity or inhomogeniety on this scale. (The full range of the color scale in Fig.~\ref{f:tem}D is 2\,nm.) Scanning electron microscope imaging also showed no sign of inhomogeniety at the 10\,nm length scale. 

Figure~\ref{f:xrr}A presents x-ray reflectivity data for two films (N2 and N3) and best-fit model curves (continuous lines). The nonlinear curve fitting analysis also yields film densities $\rho(TaN) \approx 15.9$\,g/cm$^{-3}$, consistent with bulk values. Film phase and composition were also analyzed using x-ray diffraction, shown in Fig.~\ref{f:xrr}B, which confirm the amorphous nature of the \tanx{} films studied here. Broad amorphous peaks at 36$^{\circ}$, 42$^{\circ}$, and 60$^{\circ}$ corresponding to the (111), (200), and (220) peaks of FCC $\delta-$TaN are visible in the curve for the thickest (250\,nm) sample N0. The curve for sample N1 (36~nm) barely shows the same FCC $\delta-$TaN peaks, as well as peaks at 38$^{\circ}$ and 44$^{\circ}$ from orthorhombic Ta$_2$O$_5$ (often seen for sputtered TaN$_x$ films~\cite{riekkinen03m}), again indicating the 1\,nm of tantalum oxide found previously. The most thin ($d \sim 4-5$\,nm) films show additional sharp peaks from the Si substrate, but no sign of crystalline order from the \tanx{} layer.

In summary, films studied here are amorphous, continuous, and homogeneously disordered. This is reflected in the electronic properties, which show disorder-limited length scales (such as the mean free path $\ell$) much smaller than the film thickness.

\section{Experimental details and superconductor-insulator transition}
\label{s:expt}

\begin{figure}[htb]
\centering
\includegraphics[width=1.0\columnwidth]{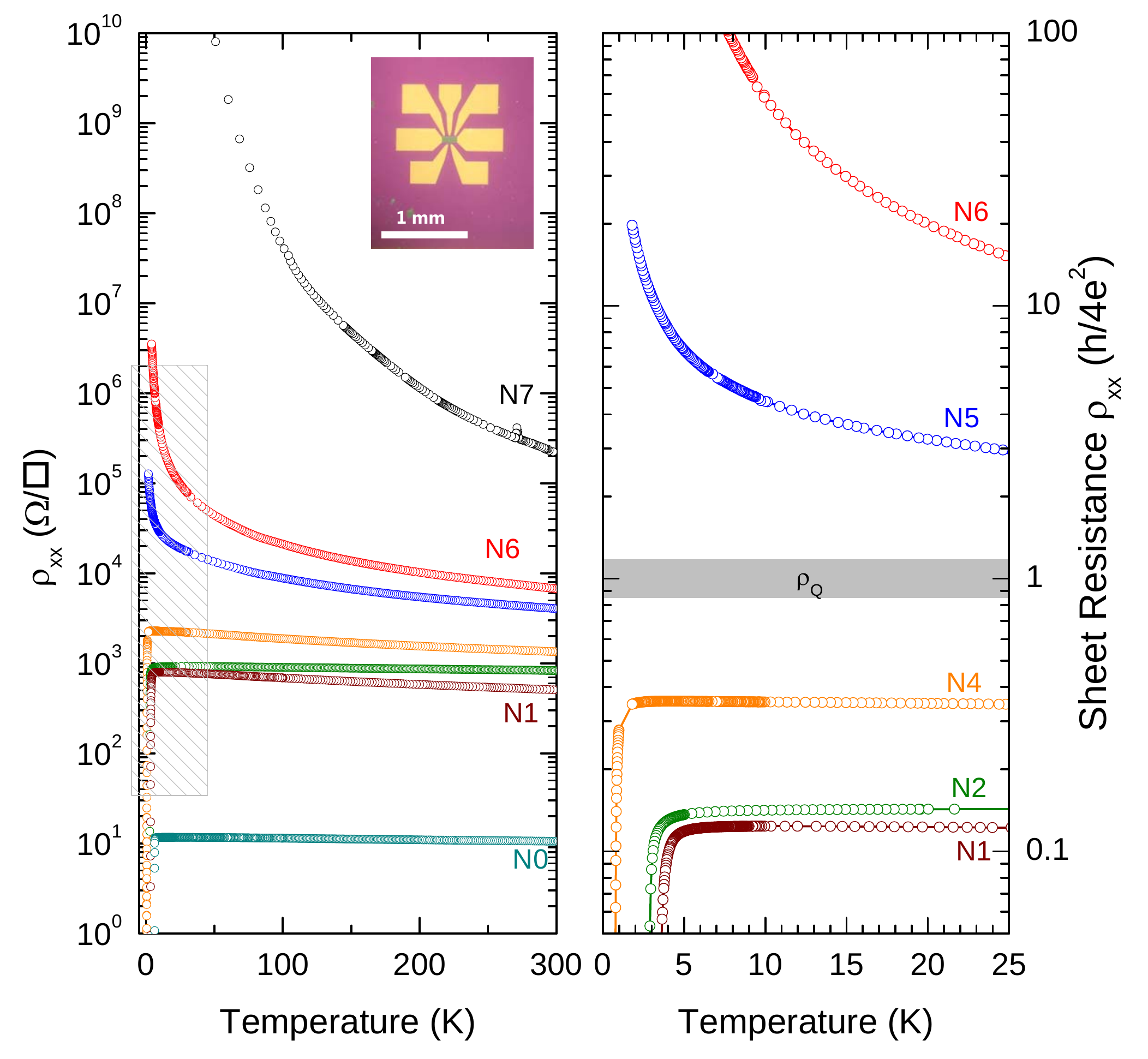}
\caption{\setlength{\baselineskip}{0.8\baselineskip} Sheet resistance versus temperature for a series of disordered Tantalum Nitride films. Disorder increases with decreasing thickness or increasing the nitrogen concentration $x$ in TaN$_x$ films. The hatched region at left is plotted in panel at right, where the horizontal gray bar indicates $\rho_{Q} = 6.45$\,k$\Omega$.}
\label{f:rvstfan}
\end{figure}

Films were patterned into Hall-bar geometry devices (see the inset of Fig.~\ref{f:rvstfan}) and measured using standard low-frequency techniques; see Ref.~\onlinecite{breznay12mt} for additional experimental details. All applied magnetic fields $H$ were perpendicular to the film plane. Care was taken to avoid electron heating during measurements below 1~K and during magnetic field ramps; all $\rxx$ and $\rxy$ measurements were confirmed to be ohmic. Reported longitudinal resistivities $\rxx$ and Hall resistivities $\rxy$ are 2D (sheet) quantities, unless explicitly noted otherwise. Table~\ref{t:params} shows electrical transport data for the eight \tanx{} films discussed here, five superconducting (N0-N4) and three insulating (N5-N7). The $T_c$ is determined where $\rxx$ has fallen to 1\% of the normal-state value, for consistency. Also shown are the 300\,K $\rxx$, Hall coefficient $R_H$, and the carrier density $n$ estimated using a single parabolic band model appropriate for these strongly disordered films.

Hall measurements show $\rxy$ linear in field and generally n-type carrier densities $\sim 10^{23}$ cm$^{-3}$, reported in Tab.~\ref{t:params}. Previous studies observed a change from n- to p- carrier type at $x\sim1.67$~\cite{yu02sm}. Bulk film resistivities are as low as $\rho_{xx} \sim 1$\,m$\Omega$-cm, depending on the stoichiometry, again comparable to films on the boundary of the stoichiometry-tuned metal-insulator transition.~\cite{yu02sm} The mean free paths for superconducting films are $\ell\sim0.1$~nm, with $k_F \ell \sim 1.0 - 3.0$ where $k_F$ is the Fermi wavevector. With the exception of the $d\sim250$\,nm thick sample N0, all films can be considered 2D in the context of superconductivity ($\xi < d$) and disorder-induced localization ($L_{\phi} < d$) effects~\cite{breznay12mt, breznay13k}.

Figure~\ref{f:rvstfan} shows the resistivity for all films as a function of temperature and illustrates the distinct superconducting and insulating ground states available to these materials as the disorder is changed. Similar behavior is observed in materials where the transition is tuned by thickness (such as in Bi~\cite{haviland1989lg}) or otherwise controlling the disorder (e.g. TiN~\cite{baturina07sb}). The separatrix between superconducting and insulating curves is $\rho_Q = h/4e^2$, highlighted in the right panel of Fig.~\ref{f:rvstfan} that shows the low-temperature region near $\rho_Q$. The superconducting $T_c$ of sputtered TaN films has been reported as high as $\sim 11$~K for thick films~\cite{kilbane75,prokhorov98,wakasugi97ts,shin01gk}; the maximum $T_c$ observed here is 6.9 K for thickest (250~nm) film studied, N0.

\begin{figure}[htb]
\centering
\includegraphics[width=1.0\columnwidth]{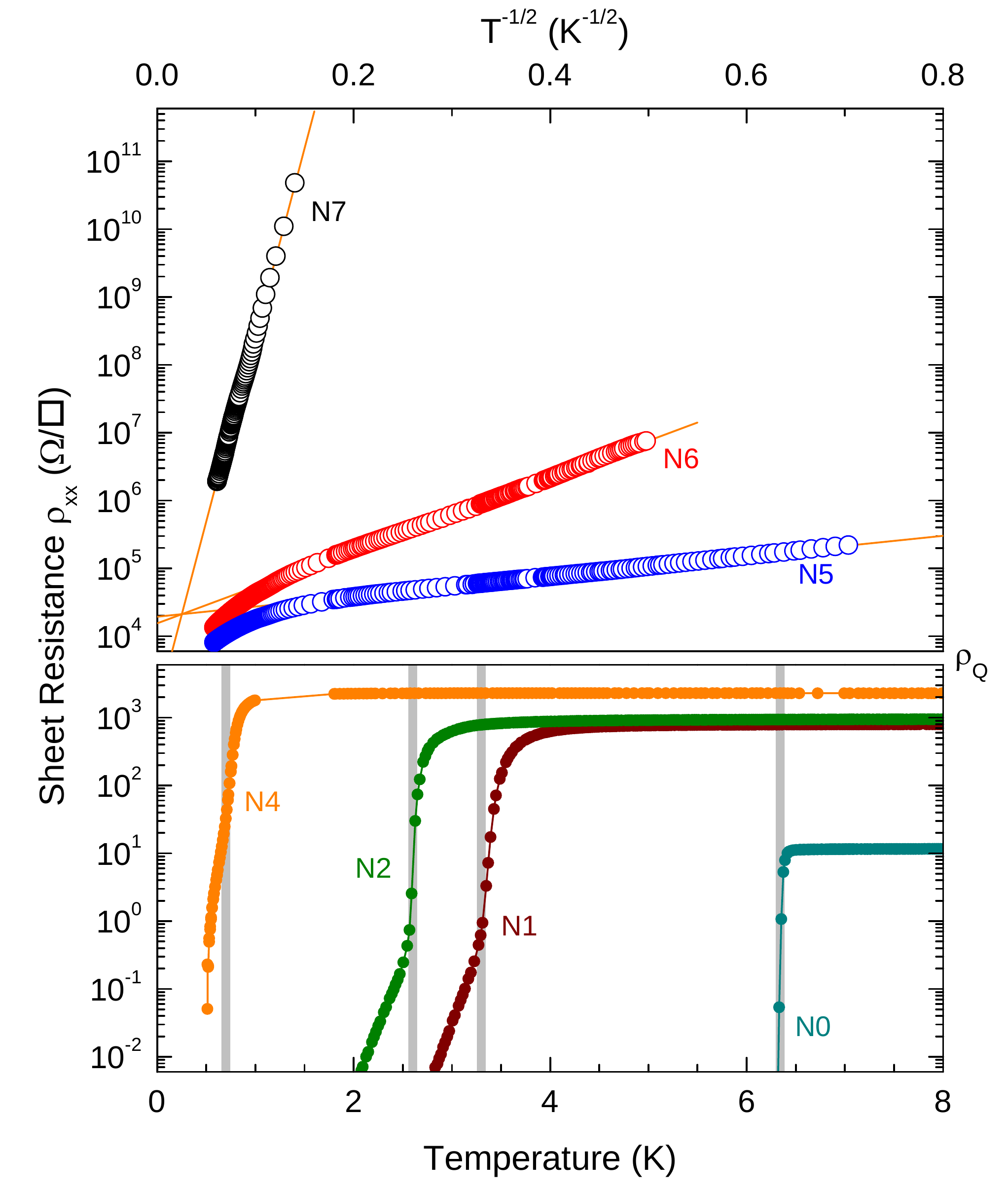}
\caption{\setlength{\baselineskip}{0.8\baselineskip} Sheet resistance $\rxx$ of insulating TaN$_x$ films (upper panel) versus $T^{-1/2}$; the linear behavior is consistent with Efros-Shklovskii~\cite{efros_1975_coulomb} hopping transport arising from a depletion of the electronic density of states due to coulomb interactions. Superconducting films (lower panel) show a crossover from 3D to 2D behavior, and finite size effects appearing as a tail below $T_c$ (gray bars); lines are guides to the eye.}
\label{f:supins}
\end{figure}

\section{Zero field ground states}
\label{s:ground}

Here we characterize the superconducting and insulating ground states of the \tanx{} films in the absence of an applied field, as the disorder is tuned via nitrogen stoichiometry. Beginning with the highest-disorder samples, Fig.~\ref{f:rvstfan} shows strong insulating behavior for samples N5-N7, with $\rxx$ increasing exponentially with temperature. The upper panel of Fig.~\ref{f:supins} shows $\rxx$(T) plotted versus $T^{-1/2}$, along with straight-line fits of the Efros-Shklovskii form:
\begin{equation}
\rxx(T) = \rho_0 \exp\left[ \left(T_0 / T \right)^{1/2} \right]
\end{equation}
where $\rho_0$ and $T_0$ are fitting parameters; the values for $T_0$ range from 50~K for sample N7 to 2~K for N5 as often observed in amorphous insulating materials.~\cite{rosenbaum_crossover_1991} The lowest measurement temperatures accessible for the insulating films are $\sim T_0$, hence we cannot conclusively establish the presence of a coulomb gap. However, both samples N6 and N7 are strongly localized insulators with clear hopping transport, and sample N5 is a marginal insulator; none of the insulating films show activated behavior characteristic of the insulating state adjacent to superconductivity in e.g. InO$_x$~\cite{sambandamurthy04ej, steiner05k}.

Less disordered films (N0-N4) with lower nitrigen content $x$ (or increased thickness) show quantum corrected metallic behavior below 20~K, with temperature-dependent conductivity contributions $\Delta\sigma(T) \sim \frac{e^2}{\pi \hbar} ln(T)$ from weak antilocalization and disorder-enhanced coulomb interactions in the normal state. Similar $\ln(T)$ temperature dependence is also observed in the normal state dependence of $\rxy(T)$. The lower panel of Fig.~\ref{f:supins} highlights the superconducting transition at $T_c$ for the samples shown, which ranges from $T_{c0} \sim 6.35$~K (sample N0) to 0.5~K (sample N4). With increasing disorder, the $T_c$ of samples N1-N4 is suppressed; this can be attributed to the effect of enhanced coulomb interactions~\cite{breznaythesis}, as demonstrated in MoGe~\cite{finkelshtein87,graybeal84b}. The increase in disorder arises both from reducing film thickness (samples N0, N1, and N2) as well as increasing nitrogen composition $x$ between samples N2, N4, and (insulating) N5. Resistive transitions are broadened due to enhanced superconducting fluctuation conductivity above $T_c$~\cite{breznay12mt}. Below $T_c$, superconducting films all show a `foot' arising from a cutoff in the diverging correlation length $\xi$ due to film inhomogeniety~\cite{benfatto_broadening_2009} or finite magnetic field~\cite{hsu92k}; further details of the superconducting properties of these films can be found in Ref.~\onlinecite{breznaythesis}.

\section{Field-tuned SIT}
\label{s:bsit}

\begin{figure}
	\centering
	\includegraphics[width=0.95\columnwidth]{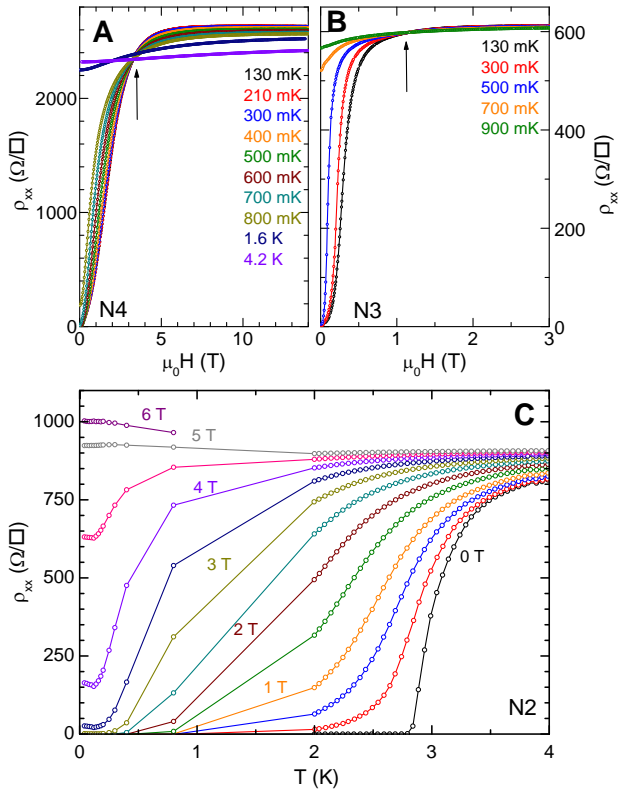}
	\caption{\setlength{\baselineskip}{0.8\baselineskip} (A-B) Magnetic field-tuned SIT in disordered \tanx{} films, evidenced by magnetoresistance isotherms for films N4 and N3. (C) Sheet resistance is plotted as a function of temperature for magnetic fields between 0 and 6 T as indicated, in equal steps. Lines are guides to the eye.}
	\label{f:bsit}
	\end{figure}

Superconductivity can be suppressed with application of critical fields $\mu_0 H_{c2} \approx 1-5$~T, revealing an insulating normal state as $T\rightarrow0$. Figure~\ref{f:bsit}A (B) shows magnetoresistance isotherms for sample N3 (N4) at temperatures near $T_c = 0.5$\,K (0.7\,K); the approximate value for $H_c$ is indicated with an arrow. For $H < H_c$ the sample is superconducting; $\rxx$ decreases with decreasing temperature. At high magnetic fields $\rxx$ shows insulating behavior, diverging as $T\rightarrow0$.

The field-tuned transition between superconducting and insulating temperature dependence is shown for sample N2 in Fig.~\ref{f:bsit}C, which plots $\rxx$ versus $T$ curves in zero field and in applied fields of up to 6~T. The low-field curves show superconducting behavior with the value of $T_c$ gradually suppressed with increasing magnetic field, the 6~T trace shows a negative slope $d\rho/dT$ to the lowest temperature available, $\sim$100~mK, with $\rho_{xx} \sim \ln(T)$ again as expected for a quantum-corrected diffusive system. This insulating behavior persists unchanged with increasing magnetic fields.

At $H_c$ there exists a temperature-independent crossing point, whose appearance is a hallmark for the SIT. In the ``dirty boson'' scenario for the SIT, the behavior near this quantum phase transition is described by the localization of Cooper paris (or proliferation of vortices) in the presence of disorder at $H_c$. Tuning via magnetic field across the putative quantum critical point at $H_c$ allows to investigate the critical scaling behavior in its vicinity.

\section{Scaling near the SIT}
\label{s:scaling}

Having identified an apparent SIT in films where disorder has already suppressed $T_c$ below the bulk value $T_{c0}$, we investigate the critical scaling behavior in the vicinity of the critical field $H_c$.~\cite{fisher90, fisher90gg} Approaching the field-tuned transition (b-SIT) there is a diverging correlation length $\xi \sim (H-H_c)^{-\nu}$ with critical exponent $\nu$. Nonzero temperatures cut off the vanishing frequency scale $\Omega \propto \xi^{-z}$, creating an additional length scale  $L_{\mathrm{th}} = (\hbar/k_B T)^{1/z}$ (here $z$ is the dynamic critical exponent). Observables such as $\rxx$ or $\rxy$ must (in 2D) be a universal function $\tilde{f}$ of the ratio $\xi/L_{\mathrm{th}}$, through
	\begin{equation}
	\rho_{xx}(H,T) = \rho_c \tilde{f}\left(\frac{H-H_c}{T^{1/z \nu}} \right)
	\label{eq:rxxscaling}
	\end{equation}
where $\rho_c$ is the critical resistance at the transition, predicted in the dirty boson scenario to be $\rho_c = \rho_Q \equiv \frac{h}{4e^2}$. Thus isotherms of $\rxx$ or $\rxy$ should collapse above and below the b-SIT when plotted versus the scaling parameter $\eta \equiv \frac{H-H_c}{T^{1/z \nu}}$. The best collapse determines the product $z \nu$.

\begin{figure}
	\centering
	\includegraphics[width=1.0\columnwidth]{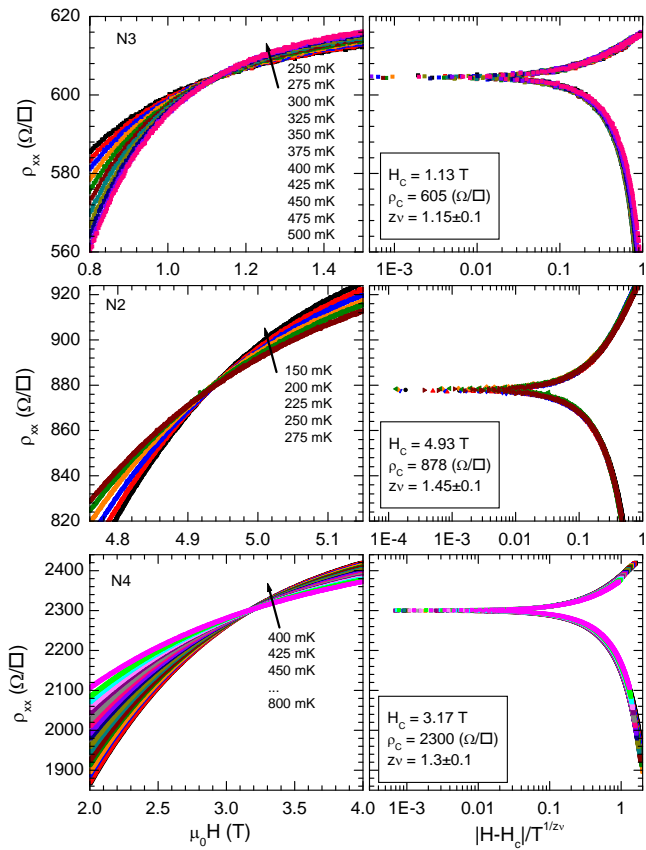}
	\caption{\setlength{\baselineskip}{0.8\baselineskip} Left panels: magnetoresistance isotherms for samples N3, N2, and N4 near the b-SIT. Right panels: resistance versus the scaling variable $\eta$ for the identical datasets plotted at left, showing excellent collapse for a range of temperatures; collected scaling parameters are shown. }
	\label{f:rxxscale}
	\end{figure}

Figure~\ref{f:rxxscale} illustrates such finite-size scaling for three \tanx{} films, N2-N4. At left the raw magnetoresistance isotherms (in steps in temperature as indicated) cross at a common critical resistance $\rho_c$, with values ranging from 0.6-2.3~k$\Omega$, all well below $\rho_Q$. At right in Fig.~\ref{f:rxxscale} are the same data versus the scaling parameter $\eta$. All three samples exhibit excellent scaling near $H_c$ over a range of temperatures (for sample N2) well below $T_c$, and for samples N3 and N4 approximately up to $T_c$, with values of $z \nu \approx 1.25$. The scaling is disrupted at the lowest temperatures; such behavior has been seen previously in e.g. MoGe~\cite{mason99k} and InO$_x$~\cite{steiner05k}. We also note that values for $H_c$ are comparable to $H_{c2}$ estimated for each of these samples (collected in Table~\ref{t:sit}). As discussed further below, these results are consistent with the ``weak insulating'' behavior seen in previous studies across a range of materials (and a range of disorder strengths)~\cite{steiner08bk}. In none of these studies, however, has similar scaling behavior been resolved in the Hall resistance $\rxy$; we consider such data in the next section.

\begin{table}[htb]
\centering
\caption{Critical \tanx{} film parameters at the SIT. $\rho_{xx}^{\text{ns}}$ is the resistance in the normal state, $T_c$ is the transition temperature,  $\rho_{c}$ and $H_c$ are the critical resistance and field at the transition, $H_{c2}$ is an estimate for the mean-field upper critical field, and $z$ and $\nu$ are critical exponents.}
\begin{ruledtabular}
\begin{tabular}{cdddddd}
     \multicolumn{1}{c}{Sample} &
     \multicolumn{1}{c}{$\rho_{xx}^{\text{n}}$} &
     \multicolumn{1}{c}{$T_c$} &
     \multicolumn{1}{c}{$\rho_{c}$} &
     \multicolumn{1}{c}{$\mu_0 H_c$}&
     \multicolumn{1}{c}{$\mu_0 H_{c2}(0)$}&
     \multicolumn{1}{c}{$z \nu$} \\
     \multicolumn{1}{c}{} &
     \multicolumn{1}{c}{(k$\Omega$)} &
     \multicolumn{1}{c}{(K)} &
     \multicolumn{1}{c}{(k$\Omega$)} &
     \multicolumn{1}{c}{(T)} &
     \multicolumn{1}{c}{(T)} &
     \multicolumn{1}{c}{} \\
     \hline
     N2 		& 0.950 & 2.75  & 0.880 & 4.93 & 5.0	& 1.45\pm0.1 \\
     N3	 		& 0.58  & 0.5   & 0.605 & 1.13 & 1.2	& 1.15\pm0.1 \\
     N4  		& 2.4	  & 0.7   & 2.30 	& 3.17 & 3.5	& 1.3\pm0.1 \\\end{tabular}
\end{ruledtabular}
\label{t:sit}
\end{table}

\section{Hall effect}
\label{s:hall}

Early studies by Paalanen et al.~\cite{paalanen92hr} found a crossing point in $\rxy$ in strongly disordered InO$_x$ films beyond $H_c$, hypothesizing that this reflected the disappearance of the bosonic insulating behavior. Initial theoretical description of the dirty bosons scenario predicted scaling in $\rxy$ in analogy with $\rxx$, with a universal value for $\rxyc$:
\begin{equation}
\rho_{xy}(H,T) = \rxyc \tilde{F}\left(\frac{H-H_c}{T^{1/z \nu}} \right).
\label{eq:hallscale}
\end{equation}
here again $\tilde{F}$ is universal function, and $\rxyc$ was predicted to have a universal value. Subsequent analyses suggested that, due to ``hidden'' particle-hole symmetry, $\rxyc$ should vanish at the transition~\cite{fisher91}; despite ongoing debate as to the nature of the SIT, there have been no further systematic studies of $\rxy$ in its vicinity. This is in part due to the challenge of measuring $\rxy$ in materials where high carrier densities $n$ lead to small signal sizes.

Early scaling theory~\cite{fisher90} also predicted that (in the case of a self-dual transition) the value of $\rxyc$ would, together with the critical longitudinal resistance $\rho_c$, be given by
\begin{equation}
(\rxyc)^2 + \rho_c^2 = \rho_Q^2.
\label{eq:rxxrxy}
\end{equation}
Given that $\rxy$ is typically much less than $\rho_Q$ for non-insulating films, this would lead to the strong prediction of $\rho_c = \rho_Q$ that has been seen in many, but certainly not all, materials. However, in the weakly disordered \tanx{} films considered here, with large $n$, $R_H \ll 1$\,$\Omega/$T, and thus $\rxy \ll 1$\,$\Omega$, Eq.~\ref{eq:rxxrxy} would suggest that $\rho_c \approx \rho_Q$. But as seen in the previous section, $\rho_c \ll \rho_Q$ for all films at the field-tuned SIT, demonstrating that the non-universal $\rho_c$ in \tanx{} cannot be explained by Eq.~\ref{eq:rxxrxy}. Alternatively, a non-universal critical Hall resistance may be estimated as $\rxyc = \rho_c \tan(\theta_H)$, where $\theta_H$ is the normal-state Hall angle.

\begin{figure}
\centering
\includegraphics[width=0.8\columnwidth]{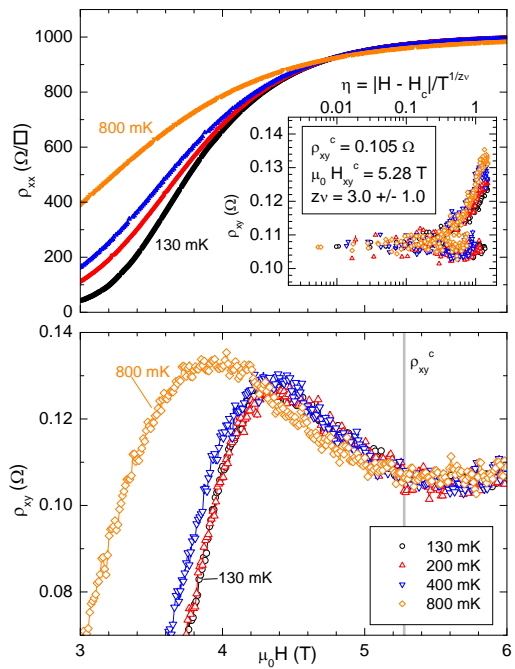}
\caption{\setlength{\baselineskip}{0.8\baselineskip} Sheet resistance $\rxx$ (upper panel) and Hall resistance $\rxy$ (lower panel) versus magnetic field for TaN$_x$ film N2. The magnetoresistance isotherms show a crossing point indicative of a field-tuned SIT; a similar crossing point can be seen in $\rxy$ (lower panel). The Hall resistance shows weak scaling in the vicinity of the crossing poing $\rxyc$ (gray bar), shown in the inset. Lines are guides to the eye.}
\label{f:hallscale}
\end{figure}

Figure~\ref{f:hallscale} shows $\rxy$ across the field-tuned SIT in film N2 (lower panel), along with simultaneously measured $\rxx$ (upper panel). The scale for $\rxy$ has been enlarged to highlight the faint crossing in $\rxy$ traces at 6.1\,T; this is above $\mu_0 H_c \approx 4.9$\,T found in the previous section. The normal state Hall angle for sample N2 is $\tan(\theta_H) = 7\times10^{-5}$ at 4.9\,T, so $\rho_c \tan(\theta_H) \approx 0.07$, below but comparable to $\rxyc = 0.105$. Following Eq.~\ref{eq:hallscale} we plot the scaled $\rxy$ in the inset of Fig.~\ref{f:hallscale}. Apparently (as with $\rxx$) the critical Hall resistance $\rxyc$ is not universal; it is well below $\rho_Q$ and the prediction $\rxyc = 0.99 \times \rho_Q$ of Eq.~\ref{eq:rxxrxy}.

\section{Discussion and summary}
\label{s:disc}

Having identified both disorder- and field-tuned SIT in \tanx{}, we can immediately address several of the key open questions surrounding this field. First, the critical disorder for the disorder suppression of superconductivity appears to be coincident with the disorder-driven (Anderson) metal-insulator transition. These TaN$_x$ films show a metal-insulator transition as a function concentration $x$, with a critical value $x_c\sim1.7$ and a critical resistivity comparable to that observed here~\cite{yu02sm}. This coincidence is seen in NbSi~\cite{crauste_destruction_2014} and NbN~\cite{chand_phase_2012}, and indicates for the SIT observed in these systems, a direct link between the appearance of (Fermionic) Anderson localization in the normal state, and the SIT.

Second, the superconductor-weak insulator transition (SWIT) observed in \tanx{} appears to be a universal phenomenon. Although the appearance of a true $T=0$ quantum phase transition is obscured by the poor scaling and unconventional metallic phase, an increasing number of materials show this phenomenology including MoGe, Ta, some InO$_x$ films, and others. In particular critical scaling (with $z \nu \sim 1.25 \pm 0.25$) is comparable to the critical exponent for classical percolation $z \nu = 4/3$, $H_c \sim H_{c2}(0)$, and $\rho_c \ll \rho_Q$~\cite{steiner08bk}. The presence of metallic ($n > 10^{22}$cm$^{-3}$) carrier densities appears to separate these materials from films of InO$_x$ that show Bose-insulator behavior. Progress towards a coherent theoretical picture for the dissipative state~\cite{davison_hydrodynamic_2016} will need to take this strong sensitivity to $n$ into account.

Third, we have found evidence for critical scaling in another observable that is complementary to $\rxx$. Here $\rxy$ appears to show a temperature-independent crossing point in both InO$_x$ and in \tanx{}; additional study is required to improve the quantitative analysis of the scaling in $\rxy$, but its value at the transition $\rxyc$ is non-universal. This contrasts with the apparent universality of $\rho_c ~ \rho_Q$ in the strong disorder limit.

And finally, disorder in these ``SWIT'' materials cannot be linked to any macroscopic granularity or inhomogeniety. Thorough investigation of the \tanx{} films showed no sign of any length scale for granular behavior in the composition or film structure; local spectroscopic probes may be necessary to study inhomogeniety length scales in the zero-temperatures phases near the SIT.

In summary, we have identified several features of the SIT in a new materials system, \tanx{}. Tuning the nitrogen content gradually suppresses superconductivity and drives a disorder-tuned SIT. Weakly insulating films with disorder-suppressed $T_c$ can be tuned via magnetic field through the SIT, and show signatures of the critical behavior arising from an underlying quantum critical point. Films with varying degrees of disorder show scaling consistent with classical percolation, with scaling disrupted at intermediate fields as $T\rightarrow0$. Both $\rxx$ and $\rxy$ show non-universal critical resistances at the SIT, and the Hall effect shows weak scaling at a critical field close to $H_c$. Thus, \tanx{} appears to represent a broad class of materials that show rich physics adjacent to the SIT that is distinct from the ``classical'' picture of disordered bosons.

\acknowledgments
We thank the staff of the Stanford Nanocharacterization Laboratory for exhaustive characterization assistance. Initial work was supported by the National Science Foundation grant NSF-DMR-9508419. This work was supported by the Department of Energy Grant DE-AC02-76SF00515.


\end{document}